\documentclass[aps,prb,twocolumn,groupedaddress,showkeys]{revtex4-1} 
\newcounter{auxFootnote}
\usepackage{fancyhdr}
\usepackage{url}

\fancypagestyle{firstpage}{
	\fancyhf{} 
	
	\fancyfoot[C]{\footnotesize The following article has been accepted by \textit{American Journal of Physics}. After it is published, it will be found at \url{https://doi.org/10.1119/5.0314108}.}
}

\usepackage{amsmath}  
\usepackage{amsfonts} 
\usepackage{graphicx} 

\usepackage[squaren,mediumqspace,amssymb]{SIunits}
\usepackage[caption=false]{subfig}

\newcommand{\s}{5\text{s}\, ^2\text{S}_{1/2}}
\newcommand{\p}{5\text{p}\, ^2\text{P}_{3/2}}

\begin{document}

\title{
	The other side of the saturation coin: an experiment to perform saturation of fluorescence spectroscopy in undergraduate laboratories.
}




\author{G. Quiroz-S\'anchez} 
\author{L. M. Hoyos-Campo}
\author{J. Flores-Mijangos} 
\author{F. Ram\'irez-Mart\'inez}\email[]{ferama@nucleares.unam.mx}
\author{J. Jim\'enez-Mier}\email[]{jimenez@nucleares.unam.mx}  
\affiliation{Instituto de Ciencias Nucleares, UNAM. Circuito Exterior, Ciudad Universitaria, 04510 Ciudad de M\'exico, M\'exico.}

\date{\today}

\begin{abstract}
	This article presents an experimental study of saturation and hyperfine pumping effects observed in fluorescence spectra when two counter-propagating beams of light interact with an atomic vapor sample. The experiment focuses on the D2 transition of atomic $^{87}$Rb by measuring changes in fluorescence emission as a room-temperature cell is exposed to two counter propagating laser beams of nearly equal intensity. The setup is comparable in complexity to saturated absorption spectroscopy experiments commonly conducted in many undergraduate laboratories but offers the advantage of a complementary view of the atom-light interaction not often found in educational laboratories. The basic theory that is needed to interpret the spectra is also presented. First, an analysis of the timescales involved in these experiments is used to identify and discriminate between different physical processes that take place during the spectroscopic measurements, thus establishing the validity of a simple model based on Einstein’s population equations. The solution of these equations is then used to calculate population surfaces that illustrate velocity-selection effects, providing an intuitive explanation of the origin of crossover and atomic resonances and the distinction between them. Both the experimental and theoretical components of this work are aimed at providing students with a thorough understanding of the interaction between electromagnetic radiation and an atomic vapor. Students with prior knowledge of laser optics, basic analog electronics, systems of coupled algebraic equations, introductory quantum mechanics, and basic atomic and statistical physics will be particularly well positioned to benefit from this presentation.
	
\end{abstract}



\maketitle 
\thispagestyle{firstpage} 

\section{Introduction} 

Numerous undergraduate laboratories routinely use narrow-band tunable diode lasers to conduct saturated absorption experiments with atomic rubidium samples. \cite{MacAdam1992,Preston1996,Razdan1999,Budker1999,Smith2004,Jacques2009,Sherlock2009}
These experiments rely on the nonlinear interaction between an atom and two electromagnetic fields.
In saturated absorption spectroscopy (SAS), a strong pump field saturates the transition, which is detected by changes in the absorption of a weak probe field \cite{Demtroeder2014a}.
Saturation effects can also be observed as Lamb dips in fluorescence emission when two counter-propagating fields of comparable intensity are used. \cite{Demtroeder2014a}
However, there are just a few examples of fluorescence saturation spectroscopy (SFS) experiments involving alkali atoms in educational laboratories.
To the best of our knowledge, only one such experiment has been reported in an undergraduate Physics Honors Project. \cite{Rowan2013}

In this paper, we present a detailed description of experiments involving the saturation of fluorescence following excitation of the D2 line in $^{87}$Rb.
The experimental setup is comparable in complexity and cost to the saturated absorption experiments typically conducted in undergraduate laboratories. However, it complements these experiments by providing a comprehensive overview of atom-light interactions.
We also review the theory of fluorescence saturation in alkali atoms, first developed by \citet{She1995}, who employed the rate equation approximation to describe the narrow peaks observed in the Doppler-broadened features of atomic sodium. Based on a comparison of the timescales of the various processes involved in this slowly varying excitation, it is demonstrated that the adiabatic elimination of transient coherent oscillations is fully justified.\cite{Meystre2021}
Therefore, a model employing Einstein's rate equations to determine the populations of the excited states in the rubidium $\p$ manifold provides an accurate description of the observed spectra.
Saturation effects arise because Doppler-shifted counter-propagating beams reduce the excited state populations compared to the simple superposition of the fluorescence produced by the two individual beams.\cite{She1995}

This article guides students through the process of adapting an standard experimental system to register more than one of the signals generated as a result of the resonant interaction of laser light and an atomic vapor and analyze the physical problem to verify the rate equation approximation validity. The set of equations describing the evolution of populations for a two level atom interacting with two resonant counter-propagating laser light components is described and next the atomic motion is incorporated by including the Doppler shifts and integrating over the Maxwell-Boltzmann distribution of velocities to calculate the saturated fluorescence profiles. Transitions between multiple atomic levels are then considered by including the hyperfine structure of $^{87}$Rb to generate spectra that can be directly compared to observations. The theoretical and experimental spectra are then contrasted and a further refinement level of the model is identified by considering the atoms magnetic structure and the selection rules related to the interaction with linearly polarized light.

In our experience, an advanced undergraduate student with skills in basic electronics and optics can be put in charge of building and testing the first version of the experiment. Once the experiment is ready, it may take up to two sessions (three hours each) to align the light beams and optimize the fluorescence detection, with an extra session to actually obtain useful spectra.
This article then aims at becoming a useful guide for a student-led project in laser science or upper-level laboratory courses exploring saturated spectra in alkali atoms.
By using Rate Equations (a standard extension of Einstein’s equations), the paper bridges the gap between introductory quantum mechanics and advanced research. 
This is achieved by breaking down the problem in clearly identified steps. The physics involved in each of these steps is well defined and clearly explained. Students then are not only guided in learning the necessary background material, but also in the process of incorporating relevant information into the analysis and interpretation of the spectra obtained in the experiment. 
This is therefore intended as a rigorous advanced laboratory guide that uses experimental evidence to teach the development, application, limits and improvement of simplified atomic models.
The intended audience of this work are then students and instructors of advanced undergraduate laboratory courses seeking to expand existing laser spectroscopy modules.

This article is organized as follows. The experimental setup, which is comparable in complexity to the standard saturated absorption setup,\cite{MacAdam1992} is described in Section \ref{sec:setup}.
The model, based on the rate equation approximation, \cite{She1995} adapted for the case of $^{87}$Rb, is discussed in Section \ref{sec:theory}.
Section \ref{sec:results} presents examples of experimental spectra alongside their comparison with theoretical results.
Finally, the conclusions are presented in Section \ref{sec:conclusions}.

\section{Experiment \label{sec:setup}}

\begin{figure}
	\centering
	\includegraphics[width=1.0 \linewidth]{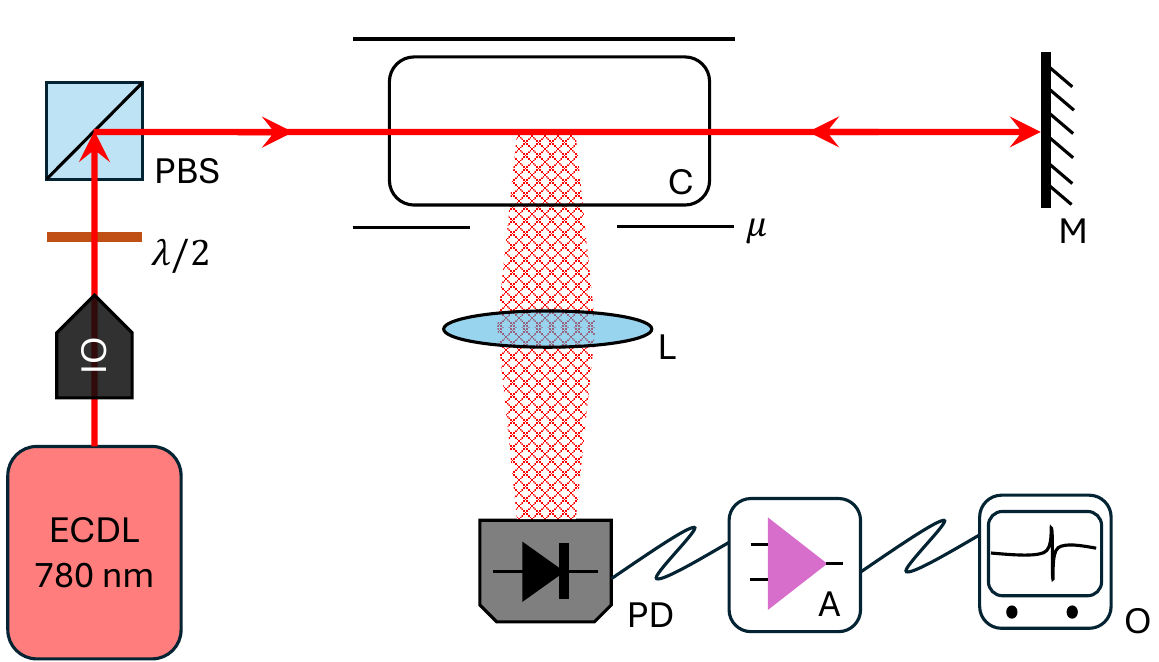}
	\caption{Experimental setup. ECDL: external cavity diode laser; IO: Faraday isolator; $\lambda/2$: half-wave plate; PBS: polarizing beam splitter; C: rubidium cell; $\mu$: $\mu$ metal layer; L: lens; M: mirror; PD: photodiode; A: amplifier; O: oscilloscope. Se the part list in the Appendix.}
	\label{fig:Setup}
\end{figure}

A diagram of the experimental setup used to record the saturated fluorescence in the D2 transition of rubidium is shown in Figure \ref{fig:Setup}. The light source was a home-built $\unit{780}{\nano\metre}$ tunable external cavity diode laser (ECDL) in the Littrow configuration.\cite{MacAdam1992, Arnold1998} The slow frequency fluctuations of this laser align with an estimated linewidth of a few $\mega\hertz$, which is comparable to the natural linewidth (FWHM) of the $\p$ excited state $\Gamma \sim  6 MHz$. The power of the laser beam was controlled using a combination of a half-wave plate ($\lambda /2$) and a polarizing beam splitter (PBS). For instance, the measurements shown in Section \ref{sec:results} were recorded with powers of up to $\sim \unit{50}{\micro\watt}$. 

The vertically polarized component of the light is reflected by the PBS into the rubidium cell (C) which contains a rubidium vapor with natural isotope abundance, that is approximately $72.2\%$ rubidium-85 ($^{85}$Rb) and $27.8\%$ rubidium-87 ($^{87}$Rb).\cite{Steck2024} The cell was maintained at room temperature and surrounded by a single layer of $\mu$-metal to attenuate the magnetic fields. After passing through the Rb cell, the beam is immediately retro-reflected by a dielectric mirror (M) at normal incidence and then traverses the cell in the opposite direction. As a result of this configuration, the two counter-propagating light components have linear and parallel polarizations in the region of interaction with the Rb vapor. During this process, students can optimize the alignment of the two beams by maximizing the sub-Doppler features observed in the saturated fluorescence spectra while changing the direction of the retroreflected beam. At this point, it is also instructional to observe the changes in the spectra as the laser power sent towards the cell is varied using the half-wave plate. It is then advised to aim for the minimum power that allows to resolve all the expected atomic and crossover lines (see section \ref{sec:results}).

Along their paths through the cell, the laser beams had an elliptical Gaussian profile with $1/e^2$ horizontal and vertical diameters of $\unit{1.93}{\milli\metre}$ and $\unit{1.32}{\milli\metre}$ ($\unit{\pm 3}{\micro \metre}$), respectively, yielding a beam area of $\unit{2\times 10^{-6}}{\metre^2}$.
Therefore, the rubidium atoms interact with collinear and counter-propagating beams with nearly equal average intensities up to $ \unit{50}{\watt\per\metre^2}$. 
The frequency of the laser was scanned  at a rate of $\unit{1}{\kilo\hertz}$ or less across the Doppler-broadened Rb D2 line resonances.  

The fluorescence due to spontaneous decay from the excited state ($\p \to \s$) is detected by a standard silicon photodiode (PD) located perpendicular to the propagation direction of the beams. 
A lens with focal length $f= 1"$ located at a distance of approximately $2"$ from both the interaction region and the photodiode (4f configuration) was used to focus fluorescence on the photodiode surface.
The photocurrent in the diode is amplified by a standard current amplifier (A). 
The output signal of the amplifier is then registered using an oscilloscope (O).
The Faraday isolator (IO) is an important element of the setup because the reflected beam is completely collinear with the input beam. 

To achieve a good signal-to-noise ratio in the fluorescence spectroscopy experiments, it is first necessary to guarantee that the light detected by the photodiode comes predominantly from the rubidium fluorescence, and that there is no stray light from the room.
The combination of a photodiode and current amplifier is highly sensitive, so background room light can easily saturate the amplifier, and it is therefore convenient to cover the cell and the detection system with opaque walls. A PVC tube or even a thick sheet of black paper was sufficient to block most of the background light.
The $\mu$-metal layer also efficiently blocks background light, but its magnetic shielding effect is only necessary if the experiment is designed to obtain more accurate results.
It is also convenient to isolate the electronics from environmental electric noise, for which the amplifier should be contained in a metallic box and a coaxial wire should always be used.
In addition, to prevent electrical noise from entering the amplifier, the wire connecting the photodiode to the amplifier should be kept as short as possible.

A list of the parts used for constructing the experimental system described here is presented in the Appendix \ref{sec:appendix}.

\section{Constructing the theoretical model\label{sec:theory}}

To analyze the saturated fluorescence spectroscopy experiments with two counter-propagating light beams of nearly equal intensities, resonantly interacting with an atomic vapor, first consider the interaction of a two-level atom with a monochromatic electromagnetic wave.
In principle, a full treatment of this problem involves studying the quantum dynamics of the density matrix of the system by solving the optical Bloch equations.\cite{Meystre2021} In this way, the time evolution of the atomic state populations, the transient oscillation of the coherent coupling between them, and the loss of these coherences can be analyzed.

\subsection{Timescales analysis}

\begin{figure}[t!]
	\centering
	\includegraphics[width=1.0 \linewidth]{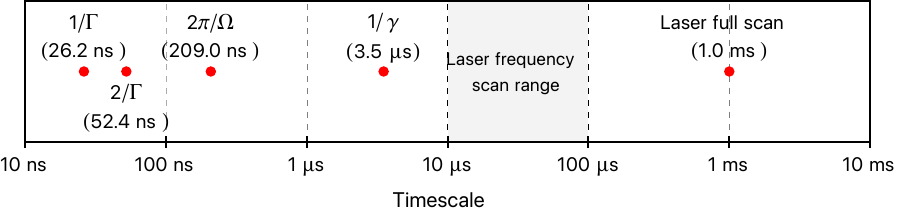}
	\caption{Characteristic timescales of atom-light interaction processes. The coherence decay time either associated only to the natural linewidth of the excited state ($2/\Gamma$) or including also the dephasing due to the laser linewidth ($1/\Gamma$) establish the timescale for the loss of coherence in the system. The timescale of coherent oscillations, the Rabi oscillation period ($2\pi/\Omega_R$), is a few times slower than the loss of coherence. The time associated with the atomic transit relaxation rate ($1/\gamma $) is respectively one and two orders of magnitude longer than the period of coherent oscillations and the coherence decay time. The full laser frequency scan is done at a rate of $\unit{\sim 1}{\kilo\hertz}$. When the entire D2 line of both Rb isotopes ($\unit{\sim 10}{\giga\hertz}$) is scanned  at this scan rate, the time for crossing through a single resonance whose width is $\Gamma$ is $\unit{\sim 10}{\micro\second}$. When a single Doppler broadened resonance is scanned (width $\unit{\sim 500}{\mega\hertz}$), the interaction times with the Doppler-free resonance is $\unit{\sim 100}{\micro\second}$.  
	}
	\label{fig:TimeScales}
\end{figure}

The choice to use the rate equation approximation is based on the assumption that the optical coherence (i.e. the off-diagonal elements of the density matrix) decays much faster than the populations of the atomic states, for which it is necessary to inspect, as illustrated in Fig. \ref{fig:TimeScales}, the characteristic timescales of the different processes involved in the measurements presented in Section \ref{sec:results}.%

The on-resonance Rabi frequency is related to the saturation parameter by $s=I/I_{\mathrm{sat}}=2 \Omega_R^2/\Gamma^2$ (see for instance Ref. \onlinecite{Meystre2021}). The saturation intensity of a given transition can be calculated using its dipole matrix element $\left|\hat{\epsilon}\cdot \mathbf{d} \right|$ by means of  $I_{\mathrm{sat}}=\frac{c\epsilon_0 \Gamma^2\hbar^2}{4\left|\hat{\epsilon}\cdot \mathbf{d} \right|^2}$, where $c$ is the speed of light in vacuum, $\epsilon_0$ is the electric constant, $\Gamma$ is the natural linewidth of the excited state and $\hbar$ is the reduced Planck constant \cite{Steck2024}. 
For instance, considering a $\pi$-polarized excitation of the $5\text{s}\, ^2\text{S}_{1/2},F=1,m_F=\pm 1 \to 5\text{p}\, ^2\text{P}_{3/2},F^\prime=1,m_F^\prime=\pm 1$ transition with a light intensity of $\unit{50}{\watt\per\metre^2}$ as indicated in Section \ref{sec:setup}, then $I_{\mathrm{sat}}^{(\pi)}= \unit{40.09}{\watt / \metre^2}$ and this gives a saturation parameter $s=I/I_s\approx 1.25$. 
In this case then a Rabi frequency of $\Omega_R=\unit{2\pi \cdot 4.8}{\mega\hertz}=\unit{3.01\times10^{7}}{\per\second}$ can be estimated, which corresponds to a Rabi period of $T_R = 209 \ \mathrm{ns}$. 

The decay of coherence considering only the natural decay rate of the excited state is $\Gamma/2$, which gives a characteristic coherence decay time of $\tau_{\rm coh}\simeq 2/\Gamma = \unit{52}{\nano\second}$. 
Furthermore, if the additional dephasing related to the laser linewidth is considered, the decay rate is approximately equal to $\Gamma$ and the coherence decay time $\tau_{\rm coh}\simeq 1/\Gamma = \unit{26}{\nano\second}$. 
Therefore, even in the best-case scenario, when $\Omega_R$ is compared solely with the natural decay rate $\Gamma$ of the excited state, $\Omega_R/\Gamma < 1$. 
This indicates that, even in this moderate saturation regime, the loss of coherence is much faster than the transient coherent oscillation in the system.%

The transit relaxation rate $\gamma=\bar{v}/\bar{d}$, as defined in ref. \onlinecite{Sagle1996}, is the rate at which an atom moving at average atomic velocity $\bar{v}$ traverses the average path length through the beam cross-section $\bar{d}$. By employing a two-dimensional Maxwell-Boltzmann distribution of velocities, the average atomic velocity is such that $\bar{v} = \sqrt{\pi k_{\text{B}} T/2 m}$, where $k_{\text{B}}$ is the Bolzmann constant, $m$ is the mass of the $^{87}$Rb atoms and the experiments were done at room temperature $T=\unit{293}{\kelvin}$.
On the other hand, as explained also in ref. \onlinecite{Sagle1996}, an average path length $\bar{d} = (\pi/4) D$ through a Gaussian beam with a full width at half maximum (FWHM) circular cross section given by $D$ is obtained by averaging all possible paths through a circle representing the laser beam cross section (see eq. 22 in ref. \onlinecite{Sagle1996}). So, 
\begin{equation}
	\gamma = \frac{\sqrt{8 k_{\text{B}} T/\pi m}}{D}.
\end{equation} 
Using the geometric mean of the $1/e^2$ beam diameters of the elliptical Gaussian profile specified in Section \ref{sec:setup}, multiplied by $\sqrt{\ln{2}/2}$ to obtain the FWHM diameter, $\gamma = \unit{284.3}{\kilo\hertz}$ and the transit time $\tau_{\rm transit}=1/\gamma \approx \unit{3.52}{\micro\second}$ is estimated. Note that this $\tau_{\rm transit}$ is one or two orders of magnitude longer than the Rabi period and coherence time.
Therefore, the coherences can be adiabatically eliminated, and the atomic dynamics can be described using classical rate equations for the populations of the ground and excited states. These equations account for optical pumping between levels with transition rates determined by the field intensity, detuning, and natural decay rate of the excited state.\cite{Meystre2021}

\subsection{Writing the rate equations}

\begin{figure}
	\centering
	\includegraphics[width=0.6  \linewidth]{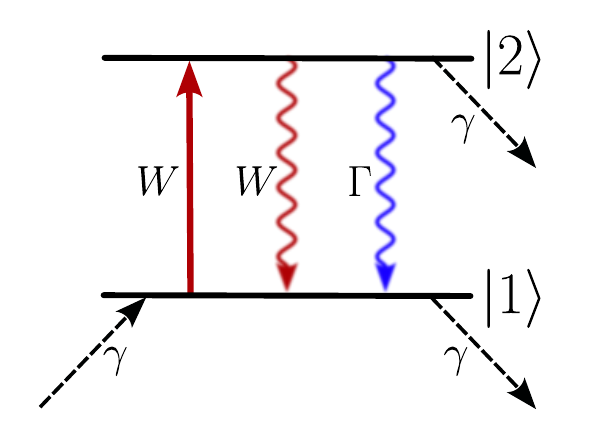}
	\caption{Two-level atom in the presence of an electromagnetic wave that produces transitions at a rate $W $. Atoms from the excited state $|2\rangle $ decay to level $|1\rangle $ by spontaneous emission at a rate $\Gamma $. Atoms in one of these two states leave the region of interaction with the wave at the rate $\gamma $, and new atoms in the lower state enter the region at the rate $\gamma $.}
	\label{fig:twolevelatom}
\end{figure}

After verifying that the typical validity conditions for the rate-equation approximation are entirely fulfilled and that it is therefore fully justified to restrict the study to this approximation, even though a narrow-band laser is used to perform the experiment, the interaction of a two-level atom with one monochromatic electromagnetic wave can be schematically represented according to the diagram shown in Fig. \ref{fig:twolevelatom}.
The electromagnetic wave couples the lower state $|1\rangle $ and upper state $|2\rangle $ inducing transitions between them.
First, absorption transitions that pump atoms in $|1\rangle $ into $|2\rangle $ occur at a rate $W(\delta)$, where $\delta = \nu - \nu_{21} $ is the detuning of the electromagnetic wave frequency $\nu $ with respect to the atomic resonance $ \nu_{21} $.
An equivalent transition rate describes the stimulated emission resulting from transition $|2\rangle \to |1\rangle $.
There are also spontaneous transitions between $|2\rangle $ and $|1\rangle $ that occur at a rate $\Gamma $, independent of the radiation frequency or its intensity.
Fig. \ref{fig:twolevelatom} also includes the transient relaxation rate $\gamma$, which is the fraction of atoms in states $|1\rangle $ or $|2\rangle$ leaving the region of interaction or entering this region in the lower state $|1\rangle $ per unit of time.

Provided these considerations, the population equations for this system are written in the following way:
\begin{subequations}
	\begin{align}
		\frac{d N_1}{dt} & = \gamma -  \gamma N_1 + W (N_2 - N_1) + \Gamma N_2 \\
		\frac{d N_2}{dt} & = -\gamma N_2 - W (N_2 - N_1) - \Gamma N_2 
	\end{align}
\end{subequations}
\noindent where $N_1$ and $N_2$ are respectively the normalized populations in states $|1\rangle $ and $|2\rangle$, such that $N_1+N_2=1$. It is important to notice that conservation of the population is built into these equations, so
\begin{equation}
	\frac{d (N_1 + N_2 )}{dt} = \gamma (1 - N_1 - N_2 ) = 0
\end{equation} 
\noindent This system of equations is readily solved in the steady state approximation ($d N_i / dt =0$) resulting in:
\begin{subequations}
	\begin{align}
		N_2 & = \frac{W}{2 W + \gamma + \Gamma}  \\
		N_1 & = 1 - N_2 
	\end{align}
\end{subequations}
\noindent The observed fluorescence results from spontaneous transitions from the upper state with an emission rate of $\phi = N_2 \Gamma $.

The dependence of the transition rate $W $ on the frequency detuning from the atomic resonance $\delta = \nu - \nu_{21} $ and on the laser beam intensity $I $ is given by
\begin{equation}
	W(\delta) = \frac{s}{\tau} \frac{1}{1+\left(4 \pi \tau \delta \right)^2}  
\end{equation}
\noindent where $\tau =1/\Gamma$ is the lifetime of the excited state, and $s = I/I_\mathrm{sat} $ is the ratio of the laser intensity to the transition saturation intensity $I_\mathrm{sat}$.\cite{She1995,QuirozSanchez2022}

\subsection{Considering atoms in motion}

Next, consider a set of atoms moving in the presence of two counter-propagating waves of the same frequency and equal intensities.
If an atom moves with a velocity component $v_z $ along the direction of propagation of the beams, it will be exposed to two Doppler-shifted waves whose angular frequencies as seen in the rest frame of the atom are
\begin{subequations}
	\begin{align}
		\nu_a = \nu   - \frac{v_z}{\lambda}   \\
		\nu_b = \nu + \frac{v_z}{\lambda}  
	\end{align}
\end{subequations}
\noindent Neglecting coherences, it is then assumed \cite{Demtroeder2014a,She1995} that in the presence of these two fields the fluorescence emission rate is obtained by making the substitution $W = W_1 + W_2 $ in the expression for the excited state population $N_2 $, where $W_1$ and $W_2 $ are the transition rates associated to each of the two waves. This results in
\begin{equation}
	\phi(\delta,v_z) = \frac{W_1+W_2}{2 (W_1+W_2) + \gamma + \Gamma} \Gamma \label{eq:fluovz}
\end{equation}

Atoms in the vapor cell follow a Maxwellian velocity distribution. That is, the probability of finding an atom with a velocity component between $v_z$ and $v_z+\mathrm{d} v_z$ along the direction of propagation of the light is given by
\begin{equation}
	f(v_z) = \sqrt{\frac{m}{2 \pi k_{\text{B}} T}}  \exp{\left[-\frac{m v^2}{2 k_{\text{B}} T}\right]} \mathrm{d}v_z \label{eq:max}
\end{equation}
The detected macroscopic fluorescence signal averages the response of atoms in all velocity groups, $v_z:\left[-\infty,+\infty \right]$. It is therefore necessary to take into account the Doppler shift experienced by each atom multiplying the fluorescence emission rate $\phi(\delta,v_z)$ in Eq. \ref{eq:fluovz} times the Maxwellian velocity distribution given in Eq. \ref{eq:max} and integrating the result over all velocities to obtain the excited state population:
\begin{equation}
	\Phi(\delta) = \sqrt{\frac{m}{2 \pi k_{\text{B}} T}} \int_{-\infty}^{\infty} \phi(\delta,v_z) \exp{\left[-\frac{m v^2}{2 k_{\text{B}} T}\right]} \mathrm{d}v_z \label{eq:fluo}
\end{equation}

\begin{figure}[t!]
	\centering
	\includegraphics[width=1.0 \linewidth]{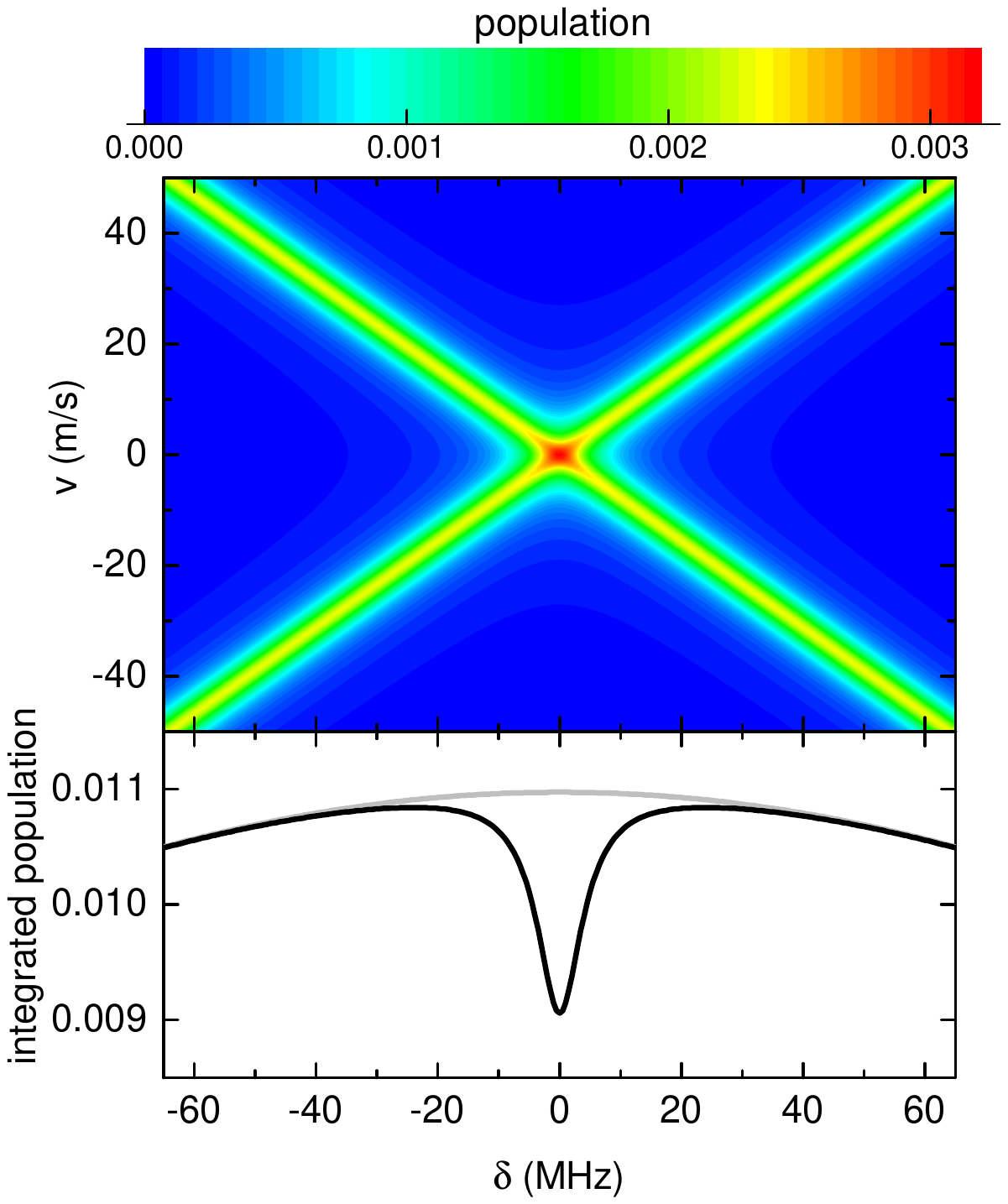}
	\caption{Upper panel: density plot of the excited state population multiplied by the velocity distribution, the integrand in Eq. \ref{eq:fluo}, as a function of the velocity and the frequency detuning. 
		Lower panel: The black line is the result of integrating the population given in Eq. \ref{eq:fluo} along constant detuning (vertical) lines, the gray line is the velocity distribution for two independent beams, with no saturation.}
	\label{fig:2LvlDensInt}
\end{figure}
A density plot of the integrand is shown in the top panel of Fig. \ref{fig:2LvlDensInt}. 
This surface has, at any given detuning, a pair of peaks that correspond to the two opposing velocity groups that are respectively Doppler shifted to resonance with each of the two counter-propagating beams. 
Equation \ref{eq:fluo} can be visualized as the integration of this surface along vertical lines that correspond to different detunings. 
The result of such integration, performed numerically, is presented in the lower panel of Fig. \ref{fig:2LvlDensInt}.
This curve indicates that the excited state population follows the Gaussian profile determined by the Maxwellian distribution (equation \ref{eq:max}), shown with a gray line, with a hole burned in the region of zero detuning.
The fluorescence obtained after the velocity integration decreases even when there is a spike in the $N_2 $ population surface because the spike maximum stays below the sum of the individual contributions of the two velocity selected groups of atoms.\cite{She1995}

\subsection{Extending the model to multilevel atoms}

\begin{figure}[htbp]
	\centering	\subfloat{\includegraphics[width=0.45\linewidth]{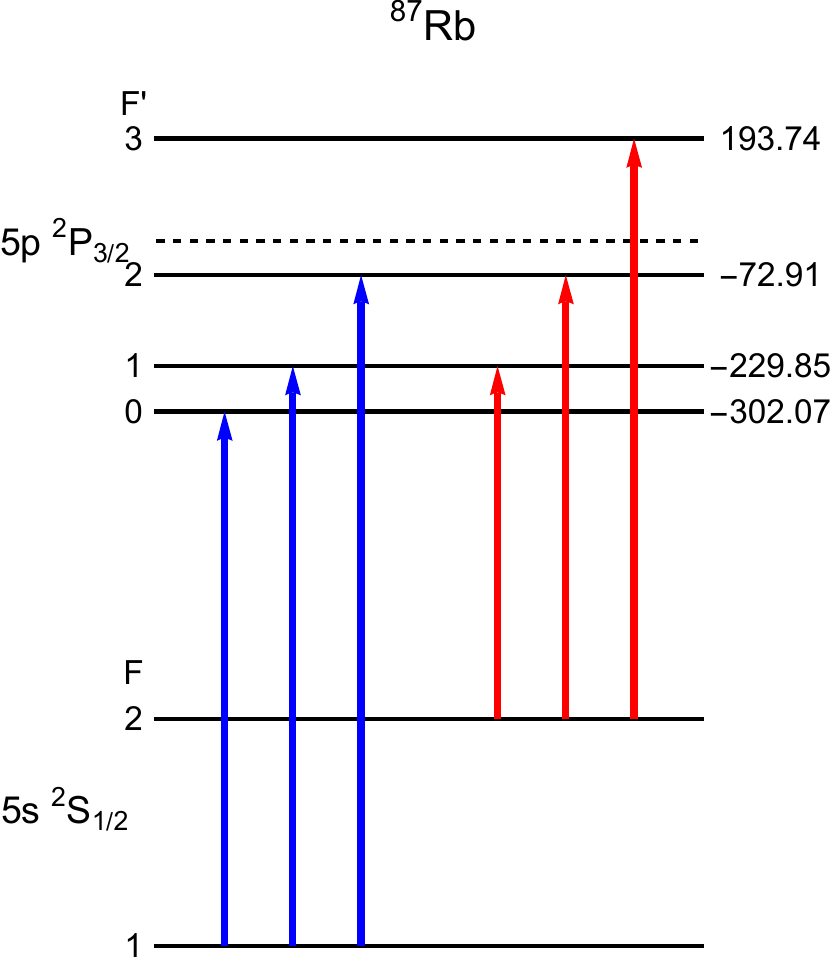}}%
	\hfill
	\subfloat{\raisebox{30pt}{\includegraphics[width=0.5\linewidth]{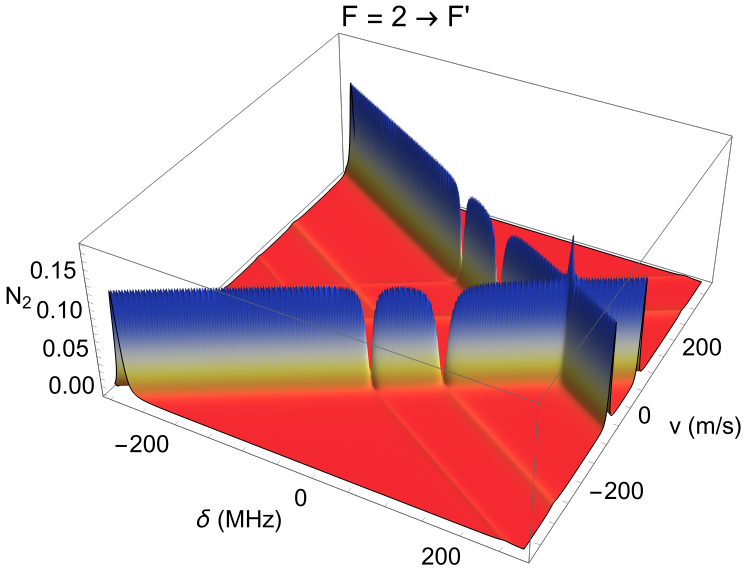}}}%
	\caption{(a) Hyperfine structure of the energy levels involved in the D2 transition in $^{87}$Rb ($I=3/2$). The three arrows to the left (blue) correspond to the electric dipole transitions from the $\s, F=1 $ hyperfine states. The three arrows to the right (red) are the transitions from the $\s, F=2 $ hyperfine state. The numbers to the right are in MHz and give the positions of the $\p $ hyperfine levels with respect to the center of gravity of the manifold, indicated by the dashed line.
		(b) Surface plot of the excited state population ($N_2$) obtained by extending the integrand in Eq. \ref{eq:fluo} to the hyperfine structure and considering the excitation starting in the $\s, F = 2 $. The zero in the detuning scale was set to the center of gravity of the $\p $ hyperfine manifold. This center of gravity is given by $\nu_{cg}= \sum_F{(2F+1)\nu_F} / \sum_F{(2F+1)} $.
	}
	\label{fig:Rb87EnLvl}
\end{figure}

This model can be extended to include the hyperfine structure of the ground and first excited states of alkali atoms.\cite{She1995,QuirozSanchez2022} For instance, Fig. \ref{fig:Rb87EnLvl}(a) shows the hyperfine energy levels involved in the D2 transition between the $\s$ and $\p$ states of $^{87}$Rb.
The $\s$ ground state splits into two hyperfine states separated by a few $\unit{}{\giga\hertz}$, each of which can be independently excited to the states in the $\p$ hyperfine manifold with a narrow bandwidth laser.
To apply the saturated fluorescence model, a set of population equations that includes detunings with respect to the pairs of states involved in each transition and the velocity of the atoms has to be solved. \cite{QuirozSanchez2022}
For example, a set of rate equations for the case of a nuclear spin $I = 3/2$, applicable to $^{23}$Na and $^{87}$Rb, is explicitly given in ref. \onlinecite{She1995}.
In the case of rubidium, if the magnetic sublevel structure is neglected, a set of five equations has to be solved: one for the active $\s$ state, i.e. the hyperfine sub-level that is resonantly coupled to the excited state by the laser, three more for the hyperfine components of the $\p$ manifold that can be excited by electric dipole transitions from this active ground state, and the last one for the hyperfine component of the $\s$ state not directly coupled by the laser and that acts as a population sink.

\subsubsection{Understanding crossover resonances}

The populations are calculated as functions of the laser detuning $\delta $ and the projection of the atom velocity along the direction of the two counter-propagating beams, $v_z $. 
A plot of the sum of the hyperfine populations $\p, F^\prime $ of $^{87}$Rb as a function of $\delta$ and $v_z $ for excitation from the $\s, F = 2 $ hyperfine state is shown in Fig. \ref{fig:Rb87EnLvl}(b).
As in the top panel of Fig. \ref{fig:2LvlDensInt}, where at any given detuning there are pairs of peaks that correspond to the two opposing velocity groups that are respectively Doppler shifted to resonance with each of the two counter-propagating beams, in the surface plot shown in Fig. \ref{fig:Rb87EnLvl}(b) these peaks appear now as pairs of ridges corresponding to the velocity groups that are resonant with the multiple transitions included in the calculation. This plot then provides a rich visual representation of the different phenomena that, when the surface is integrated, accounts for the structure found in the observed spectra.
This excited state population surface is dominated by a pair of high ridges that correspond to the excitation of the cyclic transition $\s, F= 2 \to \p, F^\prime = 3 $.
The ridge showing a positive slope with respect to $\delta$ is the result of Doppler tuning to resonance atoms with positive velocity. Conversely, atoms traveling in the negative direction are Doppler tuned to resonance along the ridge with negative slope. 
When these two ridges cross, a spike similar to the one found in Fig. \ref{fig:2LvlDensInt} appears. After integration over the velocity distribution the result is a dip in the fluorescence at this frequency detuning similar to the one found in Fig. \ref{fig:2LvlDensInt}. Therefore, this dip is caused precisely by this same saturation effect.\cite{She1995}

Parallel to these dominant ridges, two smaller ones are observed which result from Doppler tuning to the other non-cyclic hyperfine transitions, $\s, F= 2 \to \p, F^\prime = 2,1 $.
Whenever two of these ridges cross, there are strong changes in the total population of the $\p $ hyperfine states. 
Weaker spikes occur whenever smaller ridge twins with positive and negative slopes cross with each other. In these cases, the effect of saturation is barely visible after performing the integration over all velocities.

More notorious changes in the excited state populations occur whenever ridges associated to two different atomic resonances intersect, i.e. in the crossing between a strong and a weak ridge. This simultaneous Doppler tuning of a single velocity group to two different transitions is responsible for the appearance of the crossover resonances in laser spectroscopy experiments.\cite{She1995,Preston1996}
In these cases the net result is not a spike, but a drop in the $\p$ population. After integration over the velocity distribution dominant, narrow dips appear at these crossover frequencies. The origin of these dips is not saturation, but optical pumping out of the $\s, F=2$ hyperfine ground state. \cite{Smith2004}

\subsection{Including magnetic structure, light polarization and selection rules}

\begin{figure}[t!]
	\centering
	\includegraphics[width=1.0\linewidth]{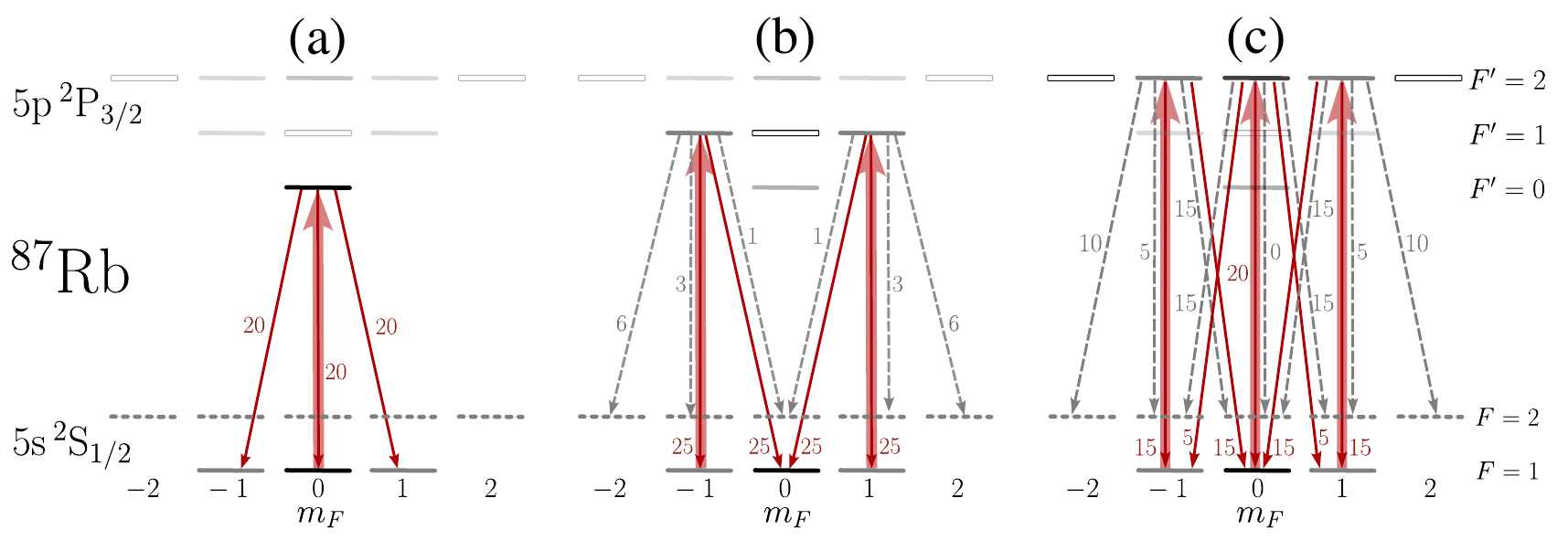}
	\caption{
		Magnetic sublevels and decay probabilities from the $\p,F^\prime=0,1,2$ states in $^{87}$Rb. The couplings generated by the $\pi$-polarized light from the $\s, F=1 $ are indicated by thick arrows. 
		Open rectangles indicate three magnetic sublevels not accessible for coupling with the $\s,F=1$ magnetic sublevels by means of $\pi$-polarized light. Dashed lines and arrows indicate respectively dark states and the decays towards them that are grouped as a single sink state. The three states with $m_F=0$ and the three pairs of sublevels with $m_F\neq 0$ are shown with solid black and gray lines respectively. 
		Decays back down to the magnetic sublevels of the lower hyperfine state, $\s, F=1 $, from each of the $F^\prime=0,1,2$ are given with red solid arrows with their relative decay probabilities. 
	}
	\label{fig:DecaysD2}
\end{figure}

The five-level calculation described so far is adequate for describing the saturation of fluorescence that follows excitation from the $\s, F = 2$ state. 
However, as it will be discussed in section \ref{sec:results}, for excitation from the lower $\s, F = 1$ hyperfine state, it is necessary to include the magnetic sublevel structure of the atoms to closely reproduce all the features observed in the experimental spectra, specifically in the one shown in Fig. \ref{fig:SatFluo87RbL}.
Fig. \ref{fig:DecaysD2} shows an schematic representation of the magnetic sublevel components that need to be taken into consideration when linearly polarized radiation excites the $\s, F=1 \to \p, F^\prime=0,1,2$ transition, including the excitation paths, decay processes and their relative probabilities. There are in total $16$ and $8$ magnetic sublevels respectively in the excited and ground states involved in this subset of possible interactions.\cite{QuirozSanchez2022} From those, the $7$ sublevels corresponding to $F^\prime=3$ are not accessible for excitation from the $F=1$ ground state and therefore are not included in these diagrams. 

The transition $\s, F=1,m_F=0 \to \p, F^\prime=1,m_F^\prime=0$ is forbidden and therefore this excited sublevel does not participate in this interaction. 
Also, the $\p,F^\prime=2,m_F^\prime=\pm 2$ magnetic sublevels are not accessible for coupling with the $\s,F=1$ magnetic sublevels by means of $\pi$-polarized light. So, only $6$ sublevels of the excited state participate in the interaction.
The five magnetic sublevels of the $\s,F=2$ hyperfine state behave as dark states that can be effectively considered as a single sink state. These sublevels and the decays towards them are shown in Fig. \ref{fig:DecaysD2} with dashed gray lines and arrows respectively. Therefore, only $3$ ground state magnetic sublevels are coupled to the excited state. 
Also, the level structure is symmetric with respect to the $m_F=0$ sublevels. The decay rates of the positive and negative $m_F$'s are identical, and the fluorescence emission spectra does not resolve decays from those pairs. Therefore, the $6$ sublevels with $m_F\neq 0$ are redundant and can be treated as only $3$ effective sublevels. This means that from the $17$ sublevels depicted in Fig. \ref{fig:DecaysD2}, it is possible to reduce the system to $7$ effective sublevels and, therefore, a set of only $7$ coupled rate equations needs to be solved.\cite{QuirozSanchez2022}

\begin{figure}[!t]
	\centering
	\subfloat{%
		\includegraphics[width=1.0\linewidth]{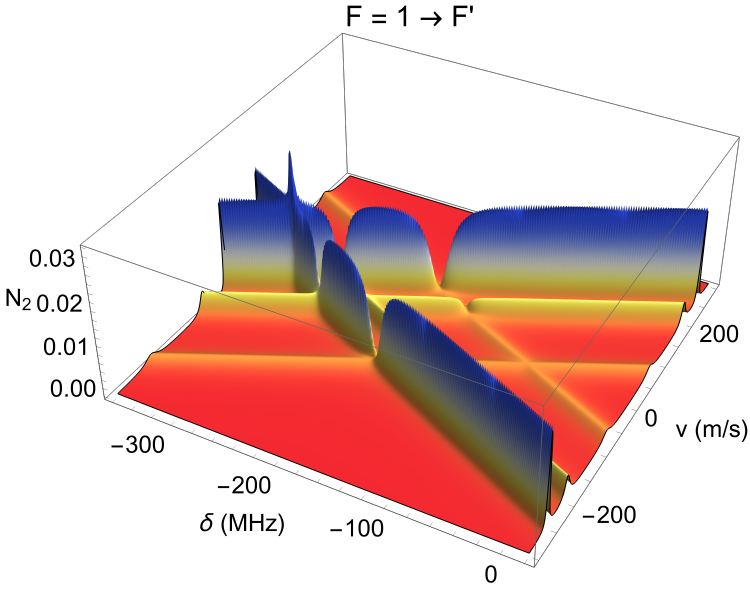}}%
	\hfill
	\subfloat{\raisebox{-10pt}{
			\includegraphics[width=1.0\linewidth]{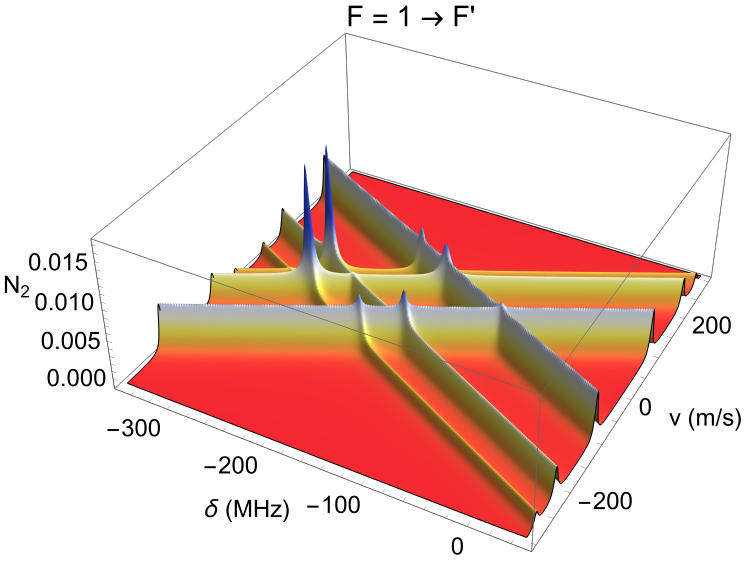}}}%
	\caption{Surface plots of the $^{87}$Rb $\p $ populations ($N_2$) obtained by extending the integrand in Eq. \ref{eq:fluo} to the hyperfine structure in the region of the $F=1 \to F^\prime $ transitions (left) calculated using hyperfine pumping (5 levels) and (right) calculated taking into account magnetic sublevel pumping (7 levels).
	}
	\label{fig:PopSurface1}
\end{figure}
The two different population surfaces shown in Fig. \ref{fig:PopSurface1} are obtained for excitation starting in the $\s, F=1 $ lower hyperfine state of the ground level, depending on whether the calculation uses five hyperfine (a) or seven magnetic sublevels (b).\cite{QuirozSanchez2022}
The population surfaces obtained with 5 levels shown in Fig. \ref{fig:PopSurface1}(a) have similar features to what was found for the $F=2 \to F^\prime $ (Fig. \ref{fig:Rb87EnLvl}): dominant ridges over the velocity groups Doppler-tuned to the cycling transitions; smaller ridges over the other atomic resonances; and population dips on the crossovers. 
On the other hand, in the population surface obtained with the seven-level calculation, given in Fig. \ref{fig:PopSurface1}(b), the ridges for the $\s, F=1 \to \p, F^\prime = 0 $ cyclic transition are no longer dominant.
There is no indication of population depletion at the crossing between ridges with opposite slopes. Instead, spikes that are more indicative of saturation effects are observed on these crossovers.

The pair of spikes at the $(0,1) $ crossover (for a detuning of $\unit{-265.96}{\mega\hertz}$) are particularly strong. 
In the diagram Fig. \ref{fig:DecaysD2}(a) it is shown that the only excitation pathway from the $\s,F=1,m_F=0$ state is to the $\p, F^\prime = 0 $ state, while diagram Fig. \ref{fig:DecaysD2}(b) shows that states with $m_F=\pm 1$ are excited to the $\p, F^\prime = 1, m_F=\pm 1$ states. Furthermore, since excitation from the $m_F=\pm 1$ states and decays towards $m_F=0$ is slightly favored $(25:20)$, there is an increase in the population of the $\s,F=1,m_F=0$ sublevel. This leads to an increment in fluorescence at the crossover. In contrast, Fig. \ref{fig:DecaysD2}(c) demonstrates that decays from $F^\prime=2$ have much higher probabilities to end up in the $\s,F=2$ dark sates and those decaying back to the $\s,F=1$ are evenly distributed among $m_F=0,\pm 1$. 
When this population surface is multiplied by the velocity distribution and integrated over the lines of constant detuning, an increase of the fluorescence is therefore found at the $(0,1) $ crossover because the sum of the heights of the pair of peaks can be higher than the Doppler background. 
For all other crossings, the effect appears as dips in the fluorescence intensity because the sum of the peaks in the rest of atomic and crossovers resonances is not as large.

\section{Saturated Fluorescence Spectra \label{sec:results}}

\begin{figure}[t!]
	\centering
	\includegraphics[width=1.0 \linewidth]{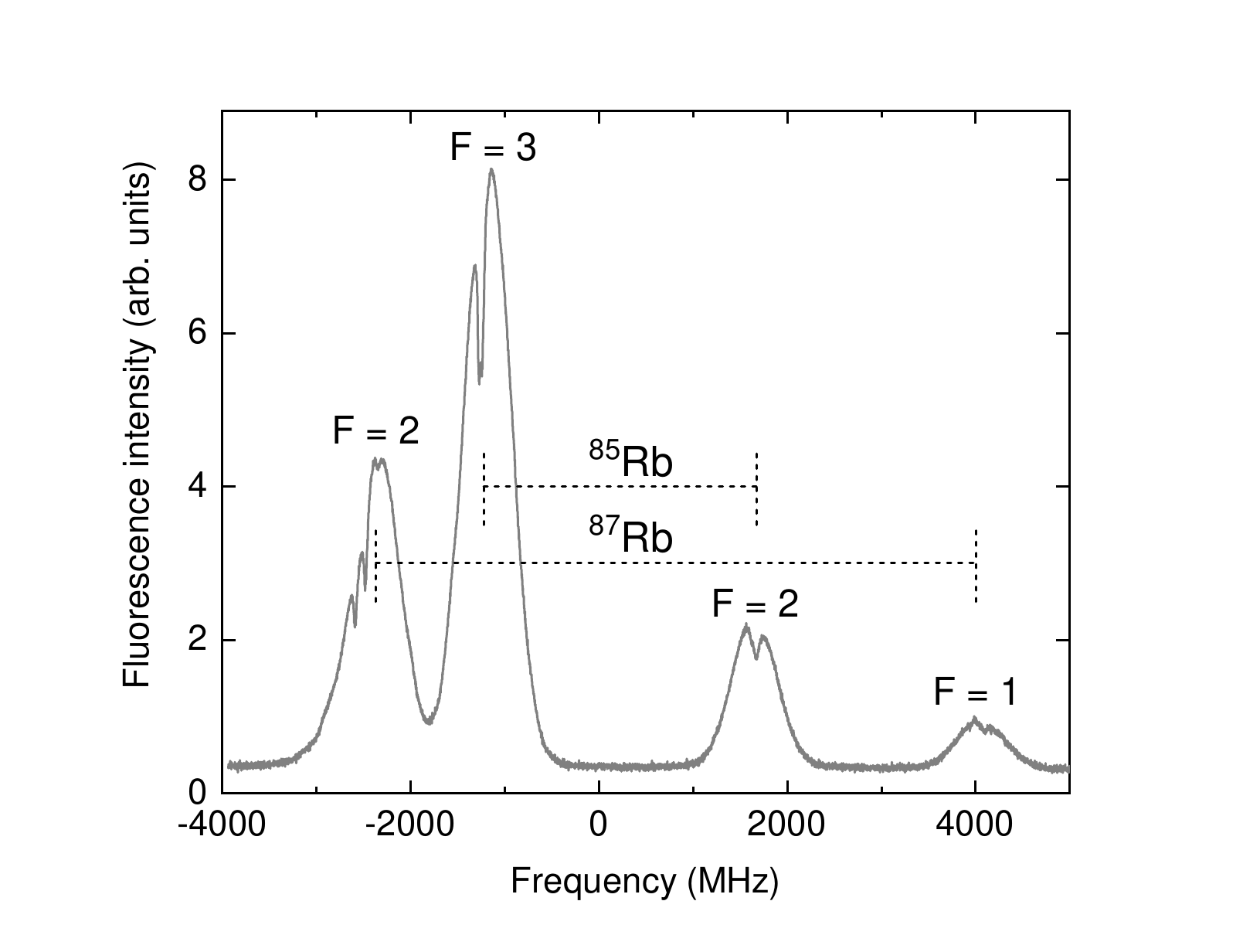}
	\caption{Saturated fluorescence spectrum for the two rubidium isotopes and for both hyperfine components of the $\s$ ground state manifold.
	}
	\label{fig:SatFluoAll4}
\end{figure}
An example of a saturated fluorescence spectrum obtained in the laboratory is shown in Fig. \ref{fig:SatFluoAll4}.
It covers the entire frequency region that comprises excitation from each of the hyperfine components of the $\s$ doublet for both $^{87}$Rb and $^{85}$Rb.
The frequency scale was calibrated with the well known $\s$ hyperfine splitting of both isotopes \cite{Steck2024}, and the zero was chosen at the common center of gravity of the $\s$ doublets.
The spectrum was recorded with two counter-propagating laser beams of near equal intensities.
It consists of four broad Gaussians, each corresponding to excitation from a specific $\s$ hyperfine component of one of the isotopes.
Structure resulting from saturation and hyperfine pumping effects is clearly visible in each of the Gaussians.

\subsection{Comparing SFS with the 5-hyperfine level model}

\begin {figure}[ht!]
\includegraphics[width = 1.0\linewidth]{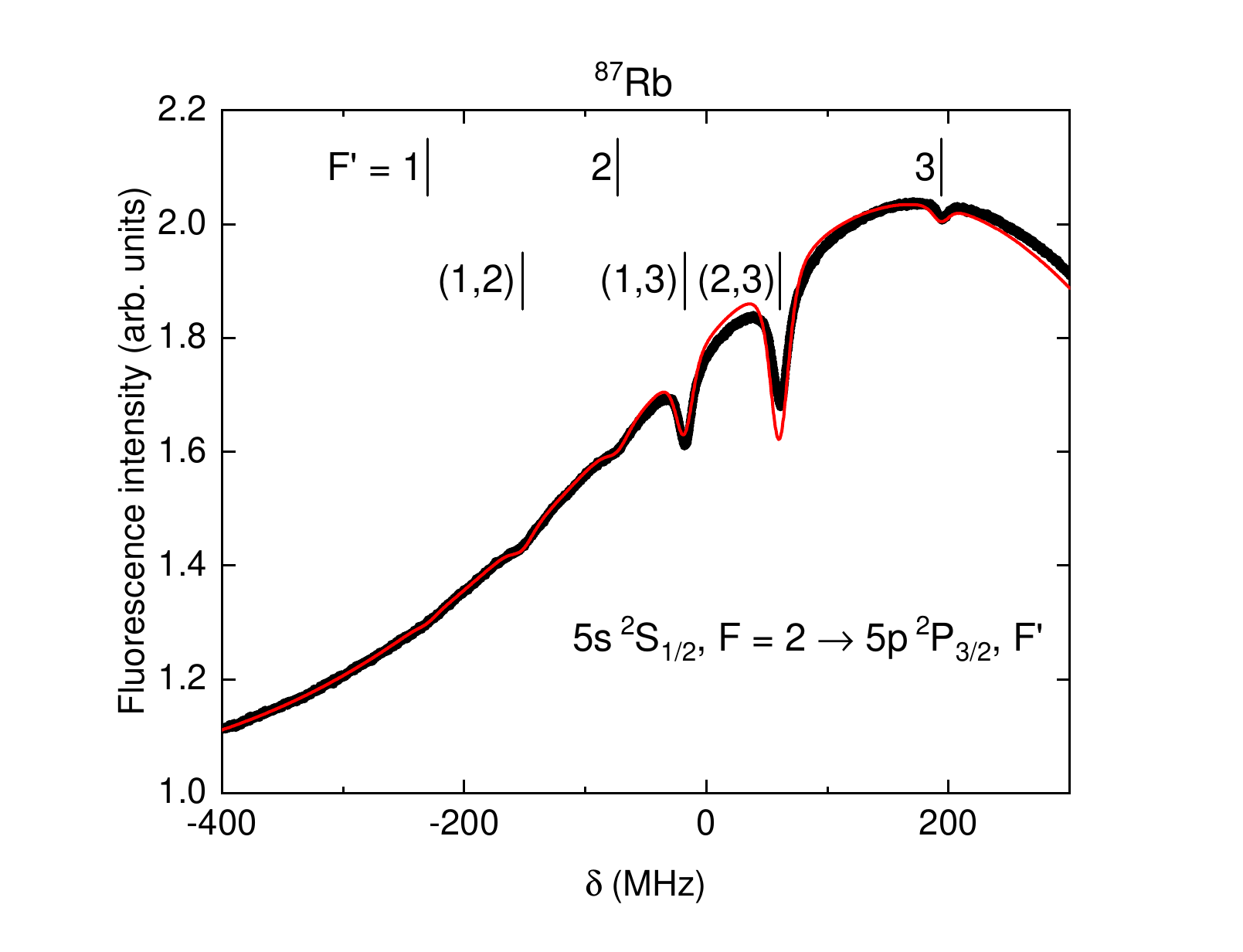}
\caption{The measured fluorescence saturation spectra in the region of the $\s, F=2 \to \p, F^\prime $ excitation in $^{87} $Rb (thick black), and comparison with the result of extending Eq. \ref{eq:fluo} to the hyperfine structure (thin red). The position of the atomic resonances and the crossovers is indicated by vertical bars.\label{fig:SatFluo87RbH}}
\end{figure}
A detailed comparison between the experimental and the calculated fluorescence saturation spectra for the region of the $\s, F=2 \to \p, F^\prime $ transitions in $^{87}$Rb is shown in Fig. \ref{fig:SatFluo87RbH}.
The zero in the frequency scale in these spectra was set to the center of gravity of the $\p$ hyperfine manifold. 
The theoretical spectrum is obtained by multiplying the population surface in Fig. \ref{fig:Rb87EnLvl}(b) by the velocity distribution (eq. 4), and integrating the result over all atomic velocities. A convolution of the resulting fluorescence with a $\unit{6}{\mega\hertz}$ (FWHM) Gaussian is also performed to take into account the laser bandwidth.
The theoretical spectrum depends on the saturation parameter $s$, which is the ratio of the laser intensity to the saturation intensity of the D2 transition.\cite{Steck2024}
In addition, a constant background and an overall amplitude factor are required in the calculation to compare the fluorescence intensity registered in the experiments with the calculated excited state population. To determine the best value for these three parameters, an optimization procedure involving the minimization of the mean square deviation of theoretically generated spectra with respect to the the experimental spectrum was performed.

The comparison between experiment and theory is very good for the $F = 2 \to F^\prime $ hyperfine manifold (Fig. \ref{fig:SatFluo87RbH}).
Both theory and experiment indicate that the most prominent features are the $(1,3)$ and $(2,3)$ crossovers that result from hyperfine pumping that removes population from the $F^\prime = 3 $ excited state in the cyclic transition.
They also agree in a weak, observable saturation peak for the $F^\prime = 3 $ atomic resonance.
The $F^\prime = 1,2 $ atomic lines, and its crossover  $(1,2)$, are very weak, but can nevertheless still be identified in the experimental spectrum and the calculation turns out to be even helpful in confirming their presence in the spectrum.

\subsection{Adding the 7-magnetic level model}
\begin {figure}[ht!]
\includegraphics[width = 1.0\linewidth]{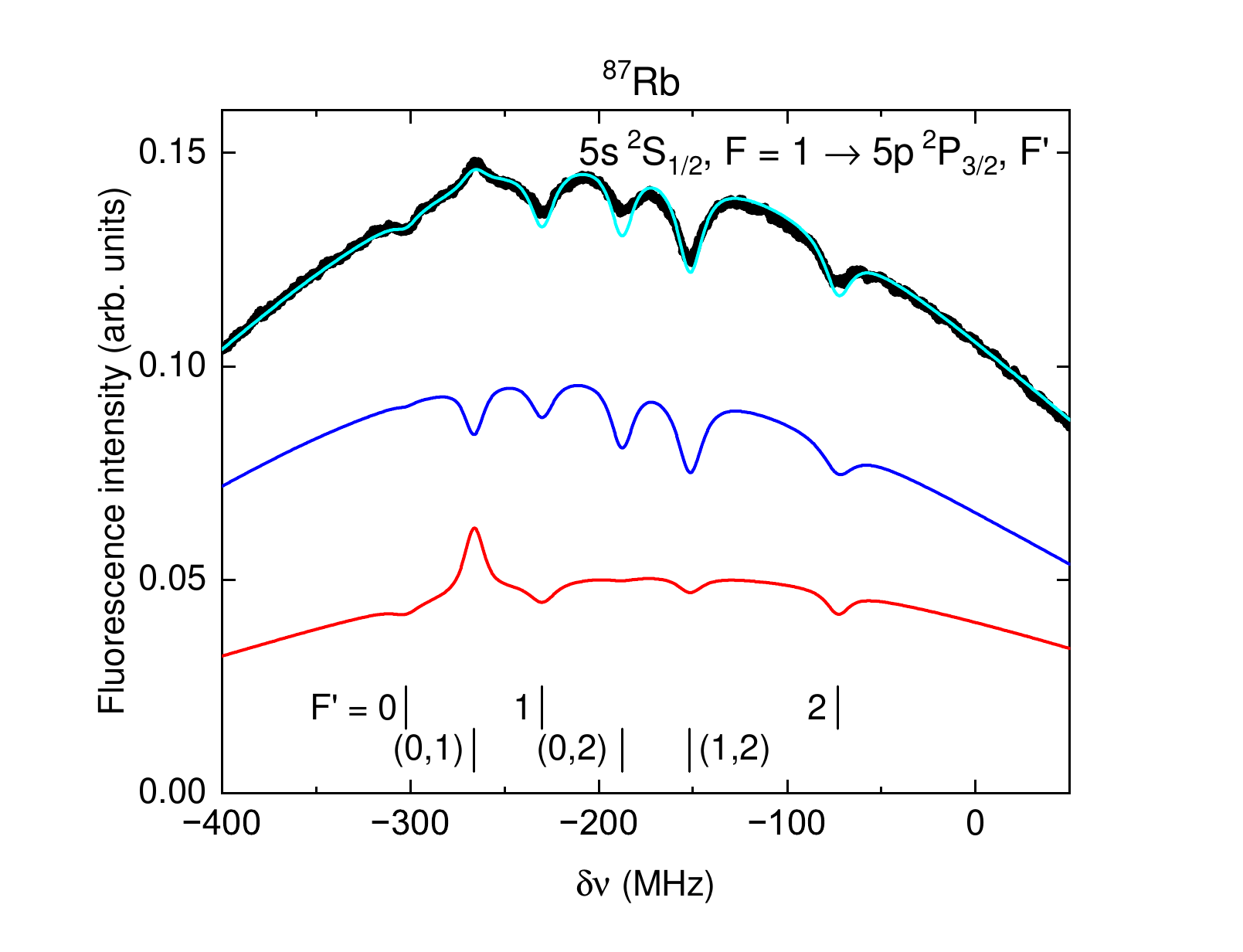}
\caption{Fluorescence saturation spectra in the region of the $\s, F=1 \to \p, F^\prime $ excitation in $^{87} $Rb, and comparison with the result of extending Eq. \ref{eq:fluo} to the hyperfine and magnetic structures. The calculated spectrum in the middle (blue) only considered the hyperfine structure, the spectrum in the bottom (red) took into account the magnetic sublevel structure. The comparison with the experiment is the sum of these two spectra (cyan). The position of the atomic resonances and the crossovers is indicated by vertical bars.\label{fig:SatFluo87RbL}}
\end{figure}
A similar comparison, now in the region of the $\s, F=1 \to \p, F^\prime $ excitation, is made in Fig. \ref{fig:SatFluo87RbL}.
In this case, two theoretical curves are included in the plot.
The curve in the middle (blue in the digital version) was obtained with the five level calculation, and it only shows dips at all the atomic resonances and crossovers.
The curve in the bottom (red online) was obtained with the seven level calculation.\cite{QuirozSanchez2022} Both curves were obtained with the same value of the saturation parameter.
The comparison between the experimental data and the sum of the two curves is given in the top curve, which is indicated by the continuous line (cyan online).
The agreement between experiment and theory is also very good in the region of the $\s, F = 1 \to \p, F^\prime $ excitations (Fig. \ref{fig:SatFluo87RbL}), provided the magnetic sublevel structure \cite{QuirozSanchez2022} is also considered. 
Indeed, in this case the best agreement is obtained by adding up the spectra calculated with the $5$ hyperfine levels and the $7$ magnetic sublevels.
The pronounced dips in the three atomic $F^\prime=0,1,2$ and the two of the crossovers $(0,2) $ and $(1,2)$ is well described by the $5$-level equations, while the enhancement in the fluorescence at the $(0,1)$ crossover only appears if the magnetic sublevel structure is included.
Interesting results are obtained by following the behavior of the saturated fluorescence spectra as a function of the laser intensity.
The general trend reported in Ref.~\onlinecite{QuirozSanchez2022} indicates that for intensities below saturation the contribution from the 5-level equations is predominant, and at larger intensities the contribution of the magnetic sublevels ($7$ equations) is the dominant one. A similar reversal from saturation to enhancement is also found in absorption spectra.\cite{r.Pappas1980, r.Schmidt1994}

\section{Conclusions.\label{sec:conclusions}}

An experiment to observe saturation effects in the fluorescence that follows the D2 excitation in atomic rubidium has been presented. 
The laboratory requirements to perform this experiment are comparable in complexity and cost to standard SAS experiments already routinely performed in many undergraduate laboratories.
Moreover, these SAS experiments can easily be modified for registering fluorescence or even to collect both signals. This guarantees experimental accessibility and low barrier to entry, besides of introducing a technique that novel uses of existing laboratory apparatus. 

The saturation effects observed in fluorescence spectra appear as sub-Doppler dips in the broad, Gaussian fluorescence emission. To analyze and interpret the measured spectra, the development of an accessible model of the population dynamics including the crossover effects has been thoroughly described. 
The physical processes that take place during the SFS measurements have been discussed and their characteristic timescales compared. This has allowed us to demonstrate why coherent oscillations can be adiabatically eliminated to provide a robust physical justification for choosing Rate Equations over Optical Bloch Equations. The procedure that has been outlined in this work clearly exemplifies the care that needs to be taken when an approximation is used in the analysis of any complex system.
A simple model based on the evolution of populations of a two-level atom has been used to explain the fluorescence saturation effect caused by two counter-propagating beams of equal intensities simultaneously interacting with an atomic vapor. 
Then, by finding the steady-state solution of these equations and including the transition rate as a function of laser detuning, the Doppler shifted resonances with the two counter-propagating laser beams have been used to integrate across the Maxwell-Boltzmann velocity distribution and obtain a numerical calculation of the Doppler broadened fluorescence profile that shows the characteristic dip caused by the saturation effect on the resonance. 

This model has been further extended to include the hyperfine sublevel structure of the states participating in the D2 transition. Calculated fluorescence saturation spectra successfully reproduce the sub-Doppler dips found in both atomic and crossover resonances. A careful revision of the calculation has shown that the origin of these crossovers is the hyperfine population pumping out of the states participating in the cyclic transition.
Furthermore, it has been shown that the surfaces obtained by calculating the number of atoms in the excited state as a function of both the laser detuning and the atomic velocity provide a rich visual demonstration of the physical processes that give rise to the various features observed in laser spectroscopy experiments, such as saturated absorption and fluorescence. These surfaces have been shown to be characterized by ridges that run along the Doppler tuned resonances for each of the counter-propagating beams, with atomic and crossover resonances appearing exactly at the detunings where these ridges cross with each other. Therefore, the spectral dips found in the saturated spectra are the result of integrating over all atomic velocities at the detunings where these ridge crossings take place. This way, students have been presented with a clear and intuitive visualization of the physical phenomena that give rise to the sub-Doppler characteristics of the observed spectra. This introduces then a powerful didactic tool for explaining and demonstrating how specific velocity groups are Doppler-shifted into resonance, making the concept of velocity-selective saturation tangible rather than purely mathematical, while remaining physically correct and accurate.

The theoretical spectra are in excellent agreement with the experimental observations in the region of the $F=2 \to F^\prime $ excitation. Nevertheless, at the $F=1 \to F^\prime $ resonances the experimental results show an enhancement of the fluorescence at the $(0,1) $ crossover. The explanation of this enhancement has been taken as an opportunity for adding one more level of complexity to the analysis. 
This not only has provided valuable insight into magnetic sublevel structure, selection rules for linearly polarized light and optical pumping effects, but also considers the full complexity of the multilevel atom structure involved in the interaction.
Although the number of atomic levels involved in the interaction became in principle unmanageable, a careful review of the problem has allowed to identify a reduced system of only seven effective levels that can again be solved without losing any relevant physical information. This accurately reproduces the crossover enhancement and gives an excellent agreement with experimental data.
To this end, it has been described how symmetry and realistic physical considerations about the system evolution can be exploited for identifying useful approximations, reducing the model complexity without loosing relevant physical information.
This then has turned into a highly educative demonstration of how a combination of intuition, insight and judicious application of selection rules elucidates the relevant physical phenomena and delivers an elegant approximation. 

In conclusion, along this article, students have been taken step-by-step through the full procedure of construction, set in operation, process of measurement, data interpretation and validation, development and refinement of a theoretical model, numerical calculation and comparison with the experiment. 
The procedure demonstrates how a theoretical model of a physics problem built from basic principles can accurately predict the observations in some situations, be limited or incomplete under different circumstances, and how additional information can be included without over-complicating the theoretical description of the system.
All this then provides a complete and exhaustive overview of the entire research procedure applied to the study of light-matter interaction, specifically in laser spectroscopy of atomic vapors. This work then provides an expansion and enhancement of pedagogical toolkits at the undergraduate and graduate levels by offering a significant instructional laboratory, data analysis and physical interpretation extension for university instructors and students alike.

\appendix*   

\section{Parts list \label{sec:appendix}}

The experimental system presented in section \ref{sec:setup} constitutes an extension to the standard SAS setup \cite{Preston1996}. The components required to build the SFS experiment are presented in Table~\ref{tab:parts}, all of which are also components normally used in standard SAS experiments. For the rubidium vapor cell and the photodiode more than one alternative is provided and the specific item utilized in this work is indicated. 

To record the fluorescence spectra, essentially it is only necessary to add a couple of low-cost elements to this setup. Firstly, to register the fluorescence radiation a photodiode with sentitivity in the near-IR region is required. For example, the  BPW34 Silicon PIN Photodiode from Vishay (\url{https://www.vishay.com/docs/81521/bpw34.pdf}) used in this work or the FDS100 Si photodiode from Thorlabs are both suitable options. Then, to build the standard current amplifier a low-power, operational amplifiers, such as the LM358 (\url{https://www.ti.com/document-viewer/LM358-N/datasheet\#abstract/snosbt30025}) used in this work needs to be considered.

\begin{table*}[!ht]
	\centering
	\caption{Parts list}
	\begin{ruledtabular}
		\begin{tabular}{@{}llc@{}}
			 \textbf{Description} & \textbf{Item} & \textbf{Supplier } \\
			\hline	
			
			\textbf{Laser light source}\\
			External cavity diode laser 780 nm  & ECDL & see \textit{e.g.} refs. \onlinecite{MacAdam1992, Arnold1998} \\  
			785 nm, Compact optical isolator & I-785-03 &  CrystaLaser \\
			\hline
			\textbf{Rubidium cell}\\
			Rubidium Wavelength Reference Cell & TT-RB-75-V-P &  Triad Technology, Inc. \\
			Rubidium Vapor Cell Pyrex\footnote{Used in this work.}\setcounter{auxFootnote}{\value{mpfootnote}} & AB-RB-P-25X75 &  Precision Glassblowing \\
			Rubidium Borosilicate Reference Cell & GC25075-RB & Thorlabs \\
			\hline
			\textbf{Optics \& optomechanics}\\
			10 mm polarizing beamsplitter cube & PBS102 & Thorlabs \\
			Kinematic prism mount & KM100PM & Thorlabs \\
			Zero order half wave plate 780 nm & WPH05M-780 & Thorlabs \\
			Rotation Mount for Ø1" (25.4 mm) Optics, 8-32 Tap & RSP1	& Thorlabs \\
			Kinematic mirror mount, with BB1-E03 & KM100-E03 & Tholabs\\
			N-BK7 plano convex lens f=25 mm\footnote{A lens tube system (\textit{e. g.} \url{https://www.thorlabs.com/half-inch-lens-tubes?tabName=Overview}) can be constructed for holding the lens and photodiode next to the Rb cell.} & LA1560-B-ML & Tholabs\\
			\hline
			\textbf{Photodiode \& amplifier}\\
			Silicon PIN photodiode\footnotemark[\value{auxFootnote}] & BPW34 & Vishay \\
			Si photodiode & FDS100 & Thorlabs \\
			Low-power operational amplifier & LM358 & Texas Instruments \\
		\end{tabular}
	\end{ruledtabular}
	\label{tab:parts}
\end{table*}

\begin{acknowledgments}
We thank J. Rangel for his help in the construction of the diode laser. This work was supported by DGAPA-UNAM, M\'exico, under projects PAPIIT Nos. IN110623 and IN118723; and by the Consejo Nacional de Ciencia y Tecnología (CONACyT) and CTIC-UNAM via the Basic Science Grant SEP-CONACyT project No. 285289, and the National Laboratories Program, Grants No. 280181, No. 293471, No. 299057 and LNC-2023-51. L.M. Hoyos-Campo thanks UNAM-DGAPA and CONACyT for the postdoctoral fellowship.
\end{acknowledgments}

\section*{Author Declarations}

The authors have no conflicts to disclose.

\subsection*{Author Contributions}

G. Quiroz-Sánchez built the experimental system and collected the experimental data, participated in the analysis of the data, and prepared the first version of the manuscript.
L. Hoyos-Campos assisted in the design of the experiment, in the collection and analysis of the experimental data and contributed to the preparation of the manuscript.
J. Flores-Mijangos assisted in the development of the experimental system and in the collection, analysis, and interpretation of the experimental data.
F. Ramírez-Martínez participated in leading the development of the project and in the analysis and interpretation of the data, and also contributed to the preparation of the manuscript.
J. Jiménez-Mier was the lead researcher and provided guidance throughout the development of the experimental system and the collection of the data. He developed the theoretical model and analysis of the experiments and contributed to the preparation of the manuscript.

\bibliography{SatFluoD2AJP.bib}

@article{r.Pappas1980,
	author = {Pappas, P. G. and Burns, M. M. and Hinshelwood, D. D. and Feld, M. S. and Murnick, D. E.},
	doi = {10.1103/PhysRevA.21.1955},
	journal = {Physical Review A},
	month = jun,
	number = {6},
	pages = {1955--1968},
	publisher = {American Physical Society},
	title = {Saturation Spectroscopy with Laser Optical Pumping in Atomic Barium},
	urldate = {2025-11-19},
	volume = {21},
	year = 1980}

@article{r.Schmidt1994,
	author = {Schmidt, O. and Knaak, K.-M. and Wynands, R. and Meschede, D.},
	copyright = {1994 Springer-Verlag},
	doi = {10.1007/BF01081167},
	issn = {1432-0649},
	journal = {Applied Physics B},
	langid = {english},
	month = aug,
	number = {2},
	pages = {167--178},
	publisher = {Springer},
	shorttitle = {Cesium Saturation Spectroscopy Revisited},
	title = {Cesium Saturation Spectroscopy Revisited: {{How}} to Reverse Peaks and Observe Narrow Resonances},
	urldate = {2025-11-19},
	volume = {59},
	year = 1994}

@article{Sagle1996,
	author = {Sagle, J. and Namiotka, R. K. and Huennekens, J.},
	doi = {10.1088/0953-4075/29/12/023},
	issn = {0953-4075},
	journal = {Journal of Physics B: Atomic, Molecular and Optical Physics},
	langid = {english},
	month = jun,
	number = {12},
	pages = {2629},
	title = {Measurement and Modelling of Intensity Dependent Absorption and Transit Relaxation on the Cesium Line},
	urldate = {2025-10-09},
	volume = {29},
	year = {1996}}

@book{Meystre2021,
	address = {Cham},
	author = {Meystre, Pierre},
	copyright = {https://www.springer.com/tdm},
	doi = {10.1007/978-3-030-76183-7},
	isbn = {978-3-030-76182-0 978-3-030-76183-7},
	langid = {english},
	publisher = {Springer International Publishing},
	series = {Graduate {{Texts}} in {{Physics}}},
	shorttitle = {Quantum {{Optics}}},
	title = {Quantum {{Optics}}: {{Taming}} the {{Quantum}}},
	urldate = {2025-10-07},
	year = {2021}}

@article{MacAdam1992,
	author = {K.B. MacAdam and A. Steinbach and C. Wieman},
	issue = {12},
	journal = {Am J Phys},
	pages = {1098-1111},
	title = {A narrow-band tunable diode laser system with grating feedback, and a saturated absorption spectrometer for {C}s and {R}b},
	volume = {60},
	year = {1992}}

@article{Jacques2009,
	author = {V Jacques and B Hingant and A Allafort and M Pigeard and J F Roch},
	doi = {10.1088/0143-0807/30/5/001},
	isbn = {0143-0807},
	issn = {0143-0807},
	issue = {5},
	journal = {Eur J Phys},
	pages = {921-934},
	title = {Nonlinear spectroscopy of rubidium: an undergraduate experiment},
	volume = {30},
	year = {2009}}

@article{Preston1996,
	author = {Daryl W. Preston},
	issue = {11},
	journal = {Am J Phys},
	pages = {1432-1436},
	title = {Doppler-free saturated absorption: Laser spectroscopy},
	volume = {64},
	year = {1996}}

@article{Razdan1999,
	author = {K. Razdan},
	doi = {10.1119/1.19131},
	issn = {00029505},
	issue = {9},
	journal = {Am J Phys},
	pages = {832},
	title = {Demonstrating optical saturation and velocity selection in rubidium vapor},
	volume = {67},
	year = {1999}}

@article{Sherlock2009,
	author = {B. E. Sherlock and I. G. Hughes},
	doi = {10.1119/1.3013197},
	issn = {00029505},
	issue = {May 2015},
	journal = {Am J Phys},
	pages = {111-115},
	title = {How weak is a weak probe in laser spectroscopy ?},
	url = {http://dx.doi.org/10.1119/1.3013197},
	volume = {111},
	year = {2009}}

@article{Smith2004,
	author = {David A. Smith and Ifan G. Hughes},
	doi = {10.1119/1.1652039},
	issn = {00029505},
	issue = {5},
	journal = {Am J Phys},
	pages = {631},
	title = {The role of hyperfine pumping in multilevel systems exhibiting saturated absorption},
	volume = {72},
	year = {2004}}

@article{She1995,
	author = {C. Y. She and J. R. Yu},
	doi = {10.1364/AO.34.001063},
	isbn = {0003-6935, 1539-4522},
	issn = {0003-6935},
	issue = {6},
	journal = {Appl Optics},
	pages = {1063},
	pmid = {21037635},
	title = {Doppler-free saturation fluorescence spectroscopy of {N}a atoms for atmospheric application},
	url = {https://www.osapublishing.org/abstract.cfm?URI=ao-34-6-1063},
	volume = {34},
	year = {1995}}

@article{Budker1999,
	author = {Dmitry Budker and Donald J Orlando and Valery Yashchuk},
	doi = {10.1119/1.19328},
	isbn = {0002-9505},
	issn = {00029505},
	issue = {7},
	journal = {Am J Phys},
	pages = {584-592},
	title = {Nonlinear laser spectroscopy and magneto-optics},
	url = {http://link.aip.org/link/?AJP/67/584/1&Agg=doi},
	volume = {67},
	year = {1999}}

@book{Demtroeder2014a,
	author = {Wolfgang Demtr{\"o}der},
	edition = {5th},
	publisher = {Springer},
	title = {Laser Spectroscopy 2. Experimental Techniques.},
	year = {2014}}

@misc{Rowan2013,
	author = {Rowan, Michael E.},
	howpublished = {Honors Thesis, Oberlin College, Oberlin, OH, USA},
	title = {Doppler-Free Saturated Fluorescence Spectroscopy of Lithium Using a Stabilized Frequency Comb},
	year = {2013}}

@article{Arnold1998,
	author = {Arnold, A S and Wilson, J S and Boshier, M G},
	journal = {Rev. Sci. Instrum.},
	number = {3},
	pages = {1236--1239},
	title = {{A simple extended-cavity diode laser}},
	volume = {69},
	year = {1998}}

@article{QuirozSanchez2022,
	author = {G. Quiroz-S{\'a}nchez and F. Ram{\'\i}rez-Mart{\'\i}nez and L.M. Hoyos-Campo and J. Flores-Mijangos and J. Jim{\'e}nez-Mier},
	doi = {https://doi.org/10.1016/j.optcom.2021.127727},
	issn = {0030-4018},
	journal = {Opt Commun},
	pages = {127727},
	title = {Saturation and optical pumping effects in the fluorescence that follows the excitation of the D2 transition in atomic rubidium},
	url = {https://www.sciencedirect.com/science/article/pii/S0030401821009305},
	volume = {508},
	year = {2022}}

@misc{Steck2024,
	author = {Daniel Adam Steck},
	title = {Rubidium 87 {D} Line Data, available online at \url{https://steck.us/alkalidata/}},
	url = {https://steck.us/alkalidata/},
	year = {date accessed February 4, 2024}}

\end{document}